\documentclass[12pt,a4paper]{article}

\usepackage{latexsym}
\usepackage{wasysym}

\title{Light Tides and the Kennedy-Thorndike experiments}

\author{Ll.\ Bel\\
\emph{Fisika Teorikoa, Euskal Herriko Unibertsitatea}, \\
\emph{P.K. 644, 48080 Bilbo, Spain}
}

\begin{document}
\maketitle

\begin{abstract}

We model the system Earth-Moon-Sun from the point of view of a frame of reference co-moving with the Earth and we derive a detailed prediction of the outcome of future Kennedy-Thorndike's type experiments to be seen as light tides.

\end{abstract}

\section*{Introduction}

Two famous experiments were meant to check an eventual anisotropy of the speed of light in vacuum depending on the direction of propagation with respect to a celestial frame of reference. The first experiment, that of Michelson-Morley \cite{MM}, as well as many others made later on, including that of  Brillet and Hall \cite{Brillet} relied on a platform that supported an interferometer and rotated on a horizontal plane. Because of this platform rotation, superimposed to the Earth rotation, these experiments tested the cosmic anisotropy of the speed of light as well as an eventual anisotropy due to any unknown local effect.

The Kennedy-Thorndike's experiment \cite{Kennedy}, as well as many others made later on and up to now \cite{Hils}-\cite{Lipa}, relies uniquely on the rotation of the Earth. We believe that these type of experiments could reveal the existence of light tides on the Earth due to the Moon and the Sun, synchronized with the fluid tides which are their counterpart. To show this, these experiments should be able to see variations of the speed of light of $1$ per $10^{17}$ parts. We understand that today some of these experiments have achieved an accuracy of $1$ per $10^{15}$ parts. We hope that the precise signature that we here derive to describe these light tides could help to fill this gap more easily.

\section{Principal transforms}

Let
\begin{equation}
\label{1.1}
d\hat s^2 =\hat g_{ij}(x^k)dx^idx^j, \quad i,j,k \dots =1,2,3
\end{equation}
be a 3-dimensional Riemannian metric defined on some manifold ${\cal V}_3$. 

We shall use the fact that in three dimensions the Ricci and the Riemann tensor are thus related:

\begin{eqnarray} 
\label{1.1.1.a}
\hat R_{ik}&=&\hat g^{jl}\hat R_{ijkl} \\
\label{1.1.1.b}
\hat R_{ijkl}&=&\hat g_{ik}\hat R_{jl}+\hat g_{jl}\hat R_{ik}
-\hat g_{il}\hat R_{jk}-\hat g_{jk}\hat R_{il}
-\frac{1}{2}\hat R(\hat g_{ik}\hat g_{jl}-\hat g_{il}\hat g_{jk})
\end{eqnarray}
where $\hat R$ is the scalar curvature.

In 1934 P.\ Walberer \cite{Walberer} proved the following theorem:

Let $\hat n_{ai}(x^k)$ be three mutually orthogonal vector fields of unit length, so that\footnote{The summation on repeated indices $a,b,c \dots =1,2,3$ is implicit}:

\begin{equation}
\label{1.2}
\hat g_{ij}=\hat n_{ai}\hat n_{aj}
\end{equation}
then there always exist locally three functions $c_a(x^k)$ such that the 3-dimen\-sion\-al metric $d\bar s^2$  with coefficients:

\begin{equation}
\label{1.3}
\bar g_{ij}=c_a^2\hat n_{ai}\hat n_{aj}
\end{equation} 
is Euclidean, i.e.:

\begin{equation}
\label{1.4}
\bar R_{ijkl}=0 \Longleftrightarrow \bar R_{ik}=0
\end{equation}
$\bar R_{ijkl}$ and $\bar R_{ik}$ being respectively the Riemann and Ricci tensors of the metric $d\bar s^2$. 

This theorem associates an Euclidean metric to any 3-dimensional one, but this association suffers from being too loose because there is an infinity of triads $\hat n_{ai}(x^k)$ that can be chosen and moreover, whatever the choice, the functions $c_a(x^k)$ depend on several arbitrary functions.

We propose here to define a tighter association following partially the lines of a preceding paper \cite{Bel96}. But it is beyond the scope of this paper to try to determine under which general circumstances the association exists or is unique.

We say that the metric $d\bar s^2$  with coefficients (\ref{1.3}) is a Principal transform of the metric (\ref{1.1}) if the following conditions are satisfied:

i) It is Euclidean: 

\begin{equation}
\label{1.5.1}
\bar R_{ijkl}=0;
\end{equation}

ii) There exist two scalar functions $A(x^k)$ and $B(x^k)$ such that:

\begin{equation}
\label{1.5.2}
\bar g_{ij}=A\hat g_{ij}+B\hat R_{ij};
\end{equation}

This implies that every eigenvector $\hat n_{ai}$ of $\hat R_{ik}$ with respect to $\hat g_{ik}$ is also an eigenvector of $\bar g_{ik}$ with respect to $\hat g_{ik}$, i.e.:

\begin{equation}
\label{1.6}
\hat R_{ik}\hat n^i_a=\rho_a\hat n_{ai} \Rightarrow \bar g_{ik}\hat n^i_a=\sigma_a\hat n_{ai}
\quad  \hat n^i=\hat g^{ik}\hat n_k
\end{equation}
and the following relationship between $\rho_a$ and $\sigma_a$:

\begin{equation}
\label{1.7}
\sigma_a=A+\rho_aB
\end{equation}
so that if one $\rho_a$ is degenerate so it is $\sigma_a$.

Let us assume that a system of coordinates exist such that in the domain of interest the metric coefficients $\hat g_{ij}$ can be written as:

\begin{equation}
\label{1.10}
\hat g_{ij}=\delta_{ij}+\hat h_{ij}
\end{equation}
where $\hat h_{ij}$, as well as its first and second derivatives, are small quantities. The corresponding principal transform will be:

\begin{equation}
\label{1.10.1}
\bar g_{ij}=\delta_{ij}+\bar h_{ij}
\end{equation}
where $\bar h_{ij}$, as well as its first and second derivatives, will be also small quantities. Coordinates that keep this forms (\ref{1.10}) and (\ref{1.10.1}) are adapted coordinates and are defined up to infinitesimal transformations:

\begin{equation}
\label{1.10.1.1}
x^k \rightarrow x^k+\zeta ^k(z^i)
\end{equation}
which induce the following transformations:

\begin{equation}
\label{1.10.1.2.a}
h_{ij}(x^k) \rightarrow 
\hat h_{ij}(x^k)+\partial_i\zeta_j(x^k)+\partial_j\zeta_i(x^k), 
\quad \zeta_i(x^k)= \delta_{ij}\zeta^j(x^k)
\end{equation}
where $h_{ij}$ is $\hat h_{ij}$ or $\bar h_{ij}$.
 
At this linear approximation $\hat R_{ijkl}$, $\hat R_{ik}$ and $\rho_a$ will be also small quantities, and the other relevant quantities that we have been mentioning can be written as:
\begin{equation}
\label{1.11}
\hat n_{ai}=n_{0ai}+\hat n_{1ai} \quad \hbox{with} \quad \delta_{ij}=n_{0ai}n_{0aj}
\end{equation}

\begin{equation}
\label{1.10.2}
\hat h_{ij}=n_{0ai}\hat n_{1aj}+\hat n_{1ai}n_{0aj}
\end{equation}

\begin{equation}
\label{1.12}
c_a^2=\sigma_a=1+\sigma_{1a} 
\end{equation}

\begin{equation}
\label{1.12.1}
A:=1+A_1, \quad B=B_0 
\end{equation}
where $n_{1ai}$, $\sigma_{1a}$  and $A_1$ are also small quantities. From the preceding expressions we derive that:

\begin{equation}
\label{1.13}
\bar g_{ij}=\hat g_{ij}+\sigma_{1a}n_{0ai}n_{0aj} 
\end{equation}
Equivalently we can write:

\begin{equation}
\label{1.13.0}
\bar h_{ij}-\hat h_{ij}=\sigma_{1a}n_{0ai}n_{0aj}
\end{equation}
or taking into account (\ref{1.5.2}):

\begin{equation}
\label{1.13.1}
\bar h_{ij}-\hat h_{ij}=A_1\delta_{ij}+B_0\hat R_{ij}
\end{equation}

Since by construction $d\bar s^2$ is Euclidean it is always possible to choose Cartesian coordinates $z^i$ for this metric. When this is done then we get the convenient formula:

\begin{equation}
\label{1.17}
\hat g_{ij}=\delta_{ij}+\hat h_{ij}(z^k) 
\end{equation}
where here:

\begin{equation}
\label{1.13.0.1}
\delta_{ij}=\bar g_{ij}, \quad \hat h_{ij}(z^k)=-\sigma_{1a}n_{0ai}n_{0aj}
\end{equation}
Or, using (\ref{1.13.1}):
\begin{equation}
\label{1.13.2}
\hat h_{ij}(z^k)=-A_1(z^k)\delta_{ij}-
B_0(z^k)\hat R_{ij} 
\end{equation}
where $A(z^k)$ and $B(z^k)$ are two scalar functions.

Let us consider a sphere with center at a point $O$ and let us use on this spherical neighborhood  a system of geodesic coordinates $y^k$ centered at $O$. We know that in this case the metric (\ref{1.1}) can be approximated up to second order by the formula:

\begin{equation}
\label{1.13.3}
\hat g_{ik}=\delta_{ik}+\hat h_{ik}(y^j), \quad \hat h_{ik}(y^j)=-\frac{1}{3}\hat R_{ijkl}y^jy^l 
\end{equation} 
the Riemann tensor being calculated at the center. Using (\ref{1.1.1.b}) this can be written:

\begin{equation}
\label{1.13.4}
\hat h_{ij}(y^k)=-\tilde A_1(y^k)\delta_{ij}-\tilde B_0(y^k)\hat R_{ij}+\frac{1}{3}(y_i\hat R_{jl}y^l+y_j\hat R_{il}y^l
+\frac{1}{2}\hat Ry_iy_j) 
\end{equation}
where:

\begin{equation}
\label{1.13.5}
\tilde A_1(y^k)=\frac{1}{3}(\hat R_{jl}-\frac{\hat R}{2}\delta_{jl})y^jy^l, \quad
\tilde B_0(y^k)=\frac{1}{3}\delta_{jl}y^jy^l
\end{equation}
Because of the last three terms in (\ref{1.13.4}) we conclude that, in this system of geodesic coordinates, $\delta_{ij}$ is not a principal transform of (\ref{1.13.3}). To achieve this, it is necessary to perform a third order infinitesimal transformation (\ref{1.10.1.1}):

\begin{equation}
\label{1.13.6}
y^i \rightarrow z^i+ \zeta^i(z^k) 
\end{equation}
to bring $\bar h_{ij}\rightarrow 0$ and: 

\begin{equation}
\label{1.13.7}
\hat h_{ik}(y^j) \rightarrow \hat h_{ik}(z^l)+\partial_i\zeta_k(z^l)+\partial_k\zeta_i(z^l) 
\end{equation}
to the convenient form (\ref{1.13.2}). These are the necessary functions $\zeta_i(z^l)$

\begin{equation}
\label{1.13.8}
\zeta_i=\left((\lambda-2/3)\hat R_{ij}\delta_{kl}-(\lambda+1/3)\hat R_{kl}\delta_{ij}
+\frac{\hat R}{12}\delta_{ij}\delta_{kl}\right)z^jz^kz^l 
\end{equation} 
The final expression of $\hat h_{ij}(z^k)$ is then:

\begin{equation}
\label{1.13.9}
\hat h_{ij}(z^k)=\left((\lambda-1)\hat R_{ij}\delta_{kl}
-(\lambda\hat R_{kl}-\frac{\hat R}{4}\delta_{kl})\delta_{ij})\right)z^kz^l 
\end{equation} 
This is as much as $\hat h_{ij}$ can be determined, in this particular case, when using Cartesian coordinates of the metric $d\bar s^2$. That is to say: there is not a unique principal transform but a one parameter family. We shall come back to this problem latter on.

\section{The Earth-Moon-Sun system}

Let us consider a stationary gravitational field referred to a frame of reference adapted to the corresponding Killing vector field. The space-time metric will be\,\footnote{We use units such that $c=G=1$.}:

\begin{equation}
\label{2.1}
ds^2=g_{\alpha\beta}(x^i)dx^\alpha dx^\beta, \quad 
\alpha,\beta,\gamma \cdots=0,1,2,3, \quad x^0=t
\end{equation}
the metric coefficients being time independent.

The physical interpretation of such gravitational fields is encoded in the following scalar and tensor potentials:

\begin{equation}
\label{2.1.1}
\xi=\sqrt{-g_{00}},\quad \varphi_i=\xi^{-2}g_{0i},
\end{equation}
and in the following quotient Riemannian metric:

\begin{equation}
\label{2.1.2}
d\hat s^2 =\hat g_{ij}(x^k)dx^idx^j, \quad 
\hat g_{ij}=g_{ij}+\xi^2\varphi_i\varphi_j 
\end{equation}
thereof the following fields are derived:

\begin{equation}
\label{2.1.3}
\Lambda_i=-\partial_i\ln\xi, \quad
\Omega_{ij}=\xi(\partial_i\varphi_j-\partial_j\varphi_i)
\end{equation}
as well as the Riemann tensor $\hat R_{ijkl}$ of the quotient metric (\ref{2.1.2}).

We consider here a very simplified, but still meaningful, description of the gravitational field of the Earth-Moon-Sun system at the linear approximation from the point of view of a 
co-rotating frame of reference co-moving with the center of the Earth. To be more precise we assume: 

i) that the coefficients of the space-time metric can be written as:

\begin{equation}
\label{2.2}
g_{\alpha\beta}=\eta_{\alpha\beta}+h_{\alpha\beta}
\end{equation}
where $h_{\alpha\beta}$ as well as its first and second derivatives are small quantities;

ii) that $h_{\alpha\beta}$ is the sum of three contributions:

\begin{equation}
\label{2.3}
h_{\alpha\beta}=h^{\earth}_{\alpha\beta}(x^i)
+h^{\leftmoon}_{\alpha\beta}(t,x^i)+h^{\sun}_{\alpha\beta}(t,x^i)
\end{equation}
where the contribution of the Earth $h^{\earth}_{\alpha\beta}$ is spherically symmetric around its center $O$ and it is time independent; and where the contributions of the Moon $h^{\leftmoon}_{\alpha\beta}$ and the Sun $h^{\sun}_{\alpha\beta}$ are those derived from their asymptotic field, their time dependence being sufficiently slow and entirely due primarily to the diurnal rotation of the Earth and secondarily to their proper motions with respect to the fixed stars. This means that for our purpose (\ref{2.1}) can be considered as static, the time evolution being considered adiabatic.

iii) that the world-line $W$ of the center of the Earth is a geodesic of (\ref{2.1})

From these assumptions it follows, using Fermi coordinates $(y^\alpha)$ in a neighborhood of $W$, that we can write:

\begin{eqnarray}
\label{2.4}
h^{\earth}_{00}(y^j)&=&-R^{\earth}_{i0k0}y^iy^k  \\
h^{\earth}_{0i}(y^j)&=&-\frac{1}{2}\Omega_{ik}y^k 
-\frac{2}{3}R^{\earth}_{ij0k}y^jy^k \\
h^{\earth}_{ik}(y^j)&=&-\frac{1}{3}R^{\earth}_{ijkl}y^jy^l
\end{eqnarray}
$R^{\earth}_{\alpha\beta\gamma\delta}$ being the contribution of the Earth to the Riemann tensor of (\ref{2.1}) on the world-line $W$, and $\Omega_{ij}$ being the constant skew-symmetric tensor describing the rotation of the Earth. We have dropped here and below in (\ref{2.7}) quadratic terms in this tensor which are irrelevant to our purpose but would be essential to discuss a Michelson-Morley experiment.

For symmetry reasons we must have:
 
\begin{equation}
\label{2.5}
R^{\earth}_{i0k0}=C\delta_{ik}, \quad R^{\earth}_{ij0k}=0, \quad
R^{\earth}_{ijkl}=K(\delta_{ik}\delta_{jl}-\delta_{il}\delta_{jk})
\end{equation}
where $C$ and $K$ are two constants that depend on the state of the center of the  Earth that we shall not need to know. None of these nor the following contributions that depend on $\Omega_{ij}$ or $R^{\earth}_{\alpha\beta\gamma\delta}$ are important to describe the diurnal or seasonal influence of the Moon and the Sun on the surface of the Earth. They are being mentioned for completeness.

The quotient metric (\ref{2.1.2}) is at this approximation:

\begin{eqnarray}
\hat g_{ik}(y^k) &=& \delta_{ik}+\hat h_{ik}(y^k) \nonumber \\
\label{2.6}
\hat h_{ik}(y^k) &=& -\frac{1}{3}(\hat R^{\earth}_{ijkl}+R^{\leftmoon}_{ijkl}(t)+R^{\sun}_{ijkl}(t))y^jy^l
\end{eqnarray}
where:

\begin{equation}
\label{2.7}
\hat R^{\earth}_{ijkl}=R^{\earth}_{ijkl}
\end{equation}
is also time independent, and where $R^{\leftmoon}_{ijkl}(t)$ and $R^{\sun}_{ijkl}(t)$ are the contributions of the gravitational field of the Moon and the Sun to the Riemann tensor on $W$ that we are going to calculate. 

Let us consider a static spherically symmetric object with mass $m$. At distances $D$ far away from its boundary its gravitational field will be weak and linear theory can be used. We know that for any system of Cartesian-like coordinates the $g_{00}$ component of the space-time metric must behave as:

\begin{equation}
\label{2.8}
g_{00}=-1+h_{00}, \quad h_{00}=\frac{2m}{D}
\end{equation}
to guarantee the correct Newtonian limit.
Taking into account that in vacuum $R_{\alpha\beta}=0$, at the appropriate approximation, we have then:

\begin{eqnarray}
\label{2.10.a}
R_{ik}&=&-R_{i0k0}+\delta^{jl}R_{ijkl}=0,  \\
\label{2.10.b}
R_{00}&=&\delta^{jl}R_{0j0l}=0
\end{eqnarray}
and therefore:

\begin{equation}
\label{2.10.1}
\hat R_{ik}= R_{i0k0}, \quad  \hat R=0
\end{equation}
where:

\begin{equation}
\label{2.10.2}
\hat R_{ik}=\delta^{jl}R_{ijkl}, \quad  \hat R= \delta^{ij}R_{ij}
\end{equation}

Taking into account that from the linearized expression of the Riemann tensor we have:

\begin{equation}
\label{2.9}
R_{i0k0}=-\frac{1}{2}\partial_{ik}h_{00}
\end{equation}
we obtain finally from (\ref{2.8}) and (\ref{2.10.1}):

\begin{equation}   
\label{2.12}
\hat R_{ik}=
-\frac{m}{D^3}(3n_in_k-\delta_{ik})
\end{equation}
where $n_i$ is the unit vector locating 
the center of the object from the point where the field is calculated. 

For $h^X_{\alpha\beta}$, where $X$ stands for $\leftmoon$ or $\sun$, we can write using (\ref{2.10.1}), the staticity assumption and obvious notations:

\begin{eqnarray}
\label{2.13}
h^X_{00}(y^k)&=&-\hat R^X_{ik}(t)y^iy^k  \\
h^X_{0i}(y^k)&=&0 \\
\hat h^X_{ik}&=&-\frac{1}{3}\hat R^X_{ijkl}(t)y^jy^l
\end{eqnarray}
where $\hat R^X_{ijkl}(t)$ could be written in terms of $\hat R^X_{ik}(t)$ using (\ref{1.1.1.b}). Therefore every $h_{\alpha\beta}$ satisfying  (\ref{2.8}) can be written in terms of $D_X$, $n_i^X$ and the geodesic coordinates $y^i$ of the metric $d\hat s^2$.

$h^X_{00}$ and $h^X_{0i}$ remain unchanged when using Cartesian coordinates $z^i$ of the metric $d\bar s^2$:

\begin{eqnarray}
\label{2.14}
h^X_{00}(z^k)&=&\frac{m^X}{D_X^3}(3n^X_in^X_j-\delta_{ij})z^iz^j \\
h^X_{0i}(z^k)&=&0 
\end{eqnarray}
This is so because the coordinate transformation (\ref{1.13.6}) that was performed to obtain (\ref{1.13.9}) did not involve the time coordinate. This equation is now

\begin{equation}
\label{2.15}
\hat h^X_{ik}=(\lambda-1)\hat R^X_{ij}\delta_{kl})z^kz^l
-\lambda\delta_{ij}\hat R^X_{kl}z^kz^l
\end{equation}

\section{Light tides and the Kennedy-Thorndike experiments}

The physical meaning of $\Lambda_i$ and $\Omega_{ij}$ in (\ref{2.1.3}) is well known as it is the meaning of (\ref{2.9}) which is responsible for the fluid tides that the Moon and the Sun induce on the surface of the Earth. On the contrary we believe that the physical meaning of $\hat g_{ij}$ and its derived objects $\hat R_{ijkl}$ or $\hat R_{ik}$ deserves to be analyzed better. They are usually interpreted as describing the geometry of space, but this metric is not Euclidean and therefore it is in conflict with the principle of free mobility of rigid bodies which is at the heart of metrology.

We propose here as we have already done before, \cite{ABMM}-\cite{Bel04} that the geometry of space it is not described by the quotient metric $d\hat s^2$ of the frame of reference but by one of its principal transform 
$d\bar s^2$. More precisely we claim that the physical distance between two points $P_1$ and $P_2$ with coordinates $x_1^i$ and $x_2^i$ as measured by an stretched unextendable thread or equivalently by a rigid rod is:

\begin{equation}
\label{3.1}
d(P_1,P_2)=\int_{P_1}^{P_2} \bar g_{ij}(x^k)dx^idx^j
\end{equation}
calculated along a geodesic of $d\bar s^2$.  If Cartesian coordinates are used then:

\begin{equation}
\label{3.2}
d(P_1,P_2)=\sqrt{\sum (z_1^i-z_2^i)^2}
\end{equation}

Let us consider a light ray that departs from a point $P$ at proper time $\tau$ of an observer at this location, bounces back at $P+dP$ and reaches the point $P$ at proper time $\tau+2d\tau$.
Light propagating along null geodesics of the space-time metric it follows that:

\begin{equation}
\label{3.3}
-d\tau^2+\hat g_{ij}(x^k)dx^idx^j=0 \quad \hbox{or} \quad \frac{d\hat s}{d\tau}=1
\end{equation}
From the usual interpretation of $d\hat s^2$ it is then concluded that light propagating along null geodesics implies that the speed of light measured along an infinitesimal round-trip circuit is always 1, using appropriate units. But this is a postulate about an interpretation of the quotient metric $d\hat s^2$ that no experiment has ever substantiated and obstructs any reasonable theory of frames of reference in General relativity.

From our point of view instead the formula (\ref{3.3}) has to be read as:

\begin{equation}
\label{3.3.1}
v(\vec e)=\frac{d\bar s}{d\tau}(\vec e)=\sqrt{\frac{\bar g_{ij}e^ie^j}{\hat g_{ij}e^ie^j}} 
\end{equation} 
where $v(\vec e)$ would be the speed of light along a direction defined by a vector $\vec e$. 

The Michelson-Morley like experiments made with an accuracy of $1$ part in $10^{13}$ and those of the Kennedy-Thorndike type made with an accuracy of $1$ part in $10^{17}$ should be able to settled this point of fundamental physics. We have discussed elsewhere, \cite{ABMM}-\cite{Bel04} these experiments  from slightly different points of view. We conclude this paper presenting the signature that we predict for the light tides that could be revealed by the Kennedy-Thorndike experiments being performed nowadays.

Using Cartesian coordinates of $d\bar s^2$ and assuming that $\vec e$ is of unit length with respect to this metric:

\begin{equation}
\label{3.3.2}
\bar g_{ij}=\delta_{ij}, \quad \delta_{ij}e^ie^j=1  
\end{equation} 
(\ref{3.3.1}) becomes at the corresponding approximation:

\begin{equation}
\label{3.3.3}
v(\vec e)=1-\frac{1}{2}\hat h_{ij}(z^k)e^ie^j  
\end{equation}
and using (\ref{1.13.9}) we obtain:

\begin{equation}
\label{3.5}
v(\vec e)=1-\frac{1}{2}\left((\lambda-1)\hat R_{ij}e^ie^j
-\frac{1}{2}(\lambda\hat R_{kl}a^ka^l-\frac{1}{4}\hat R^{\earth})\right)r^2
\end{equation}
where $r$ is the radius of the Earth, $z^i\equiv a^ir$ the coordinates of the location where the experiment is being made and where from (\ref{2.12}) we have:

\begin{equation}
\label{3.6}
\hat R_{ik}=\hat R^{\earth}_{ik}-\frac{m^{\leftmoon}}{D_{\leftmoon}^3}(3n^{\leftmoon}_in^{\leftmoon}_k-\delta_{ik})
-\frac{m^{\sun}}{D_{\sun}^3}(3n^{\sun}_in^{\sun}_k-\delta_{ik})
\end{equation}
where $\hat R^{\earth}_{ik}$ is the Earth contribution to the Ricci tensor of the quotient metric, and:

\begin{equation}
\label{3.7}
\hat R=\hat R^{\earth}
\end{equation}  
is the contribution to the scalar curvature. $\hat R^{\leftmoon}=0$ and $\hat R^{\sun}=0$ as can be seen from $\hat R^X_{ik}$ above.

In a Michelson-Morley experiment $e^k$ is a direction that rotates on a horizontal plane while in a Kennedy-Thorndike experiment $e^k$ is kept fixed. For this reason when the signal is compared for two different configurations of the triad $(a^i, n^{\leftmoon}_j, n^{\sun}_k)$ the first term in (\ref{3.6}) is irrelevant because it is time-independent, and therefore the signal should be correlated with the familiar fluid tides due to the Moon and the Sun.

As we see from (\ref{3.5}), $v(\vec e)$  depends in general on the free parameter $\lambda$ but in some cases it does not. For example if 
$\vec e$ is colinear with $\vec a$ (direction of propagation of light in the vertical direction) then we have:

\begin{equation}
\label{3.8}
v(\vec e)=v^{\earth}(\vec e)-\frac{1}{4}\frac{m^{\leftmoon}}{D^3_{\leftmoon}}r^2
(3(\vec n^{\leftmoon}\cdot\vec e)^2-1)-\frac{1}{4}\frac{m^{\sun}}{D^3_{\sun}}r^2
(3(\vec n^{\sun}\cdot\vec e)^2-1)
\end{equation} 
$v^{\earth}(\vec e)$ being the constant Earth contribution.

Other special instantaneous configuration could be considered, e.g. those with:

\begin{equation}
\label{3.9}
\vec n^{{\leftmoon}}\cdot\vec e=\vec n^{{\leftmoon}}\cdot\vec a \quad \hbox{and} \quad
\vec n^{\sun}\cdot\vec e=\vec n^{\sun}\cdot\vec a  
\end{equation} 
but these would be destroyed as time elapses. This makes particularly interesting the formula (\ref{3.8}) to test the theory independently of the value of $\lambda$.

The order of magnitude of the time-dependent terms is in any case given by:

\begin{equation}
\label{3.10}
\frac{m^{{\leftmoon}}r^2}{D_{{\leftmoon}}^3}=5\times 10^{-17}, \quad  \frac{m^{\sun} r^2}{D_{\sun}^3}=2\times 10^{-17}   
\end{equation}
 

\section{\hspace*{-.2em}The harmonic and quo-harmonic conditions}

The definition that we gave in Sect.~1 of Principal transform $d\bar s^2$ of a 3-dimensional Riemannian metric led us in the case of interest we discussed in Sect.~2 to a one parameter family of principal transforms and in Sect.~3 it led us to (\ref{3.5}) which still contains in general the free parameter.

Although this latter formula is already a neat signature for the predicted light tides and (\ref{3.8}) might already be useful it is somewhat unsatisfactory to keep the arbitrariness that this parameter represents.

From a geometrical point of view what this means is that the association between $d\hat s^2$ and $d\bar s^2$ defined in Sect.~1 is still too general and it is useful in many cases of interest to consider a supplementary condition to tighter the association. In \cite{Bel96} we chose the condition:

\begin{equation}
\label{A.8}
(\hat\Gamma^i_{jk}-\bar\Gamma^i_{jk})\hat g^{jk}=0
\end{equation}     
where $\hat\Gamma^i_{jk}$ and $\bar\Gamma^i_{jk}$ are respectively the Christoffel connection symbols of $d\hat s^2$ and $d\bar s^2$. Let us emphasize that this is, as (\ref{1.4}), a tensor equation and it is a most natural one in the context described there. We call this condition the quo-harmonic condition.

On the other hand in other contexts, in which $d\hat s^2$ is a metric derived from a more general structure, other choices might be considered. Such is the case in the context of Sect.~3 where $d\hat s^2$ is derived from a 4-dimensional space-time metric $ds^2$ taking a quotient with the time-like vector field of the corresponding frame of reference. In this case it is also a natural choice to use as supplementary condition:

\begin{equation}
\label{A.9}
(\hat\Gamma^i_{jk}-\bar\Gamma^i_{jk})\hat g^{jk}=\hat g^{ik}\Lambda_k
\end{equation}
$\Lambda^k$ being the Newtonian field of ${\cal V}_3$ defined in (\ref{2.1.3}). This is a choice of coordinates that is often made in General relativity and it is known as the harmonic condition.

Since by construction $d\bar s^2$ is Euclidean it is always possible to choose Cartesian coordinates for this metric. When this is done then (\ref{A.8}) becomes:

\begin{equation}
\label{A.8.1}
\hat\Gamma^i_{jk}\hat g^{jk}=0 \Longleftrightarrow \hat\Delta z^i=0 
\end{equation}
where $\hat\Delta$ is the Laplacian of the metric $d\hat s^2$.

And (\ref{A.9}) becomes:
\begin{equation}
\label{A.9.1}
\hat\Gamma^i_{\alpha\beta}\hat g^{\alpha\beta}=0 \Longleftrightarrow \Delta z^i=0
\end{equation}
where $\Delta$ is the Laplacian of the space-time metric $ds^2$.

Using Cartesian coordinates of the metric $d\bar s^2$ the condition (\ref{A.8.1}) becomes at the corresponding linear approximation:

\begin{equation}
\label{A.14}
\partial_i\hat h^i_j-\frac{1}{2}\partial_j\hat h=0  
\end{equation}
where:

\begin{equation}
\label{A.15}
\hat h^i_j=\delta^{ik}\hat h_{jk} \quad \hbox{and} 
\quad \hat h=\delta^{ij}\hat h_{ij} 
\end{equation}
and requires $\lambda$ to be:

\begin{equation}
\label{A.17}
\lambda=2/3 
\end{equation}
in (\ref{1.13.9}).

Similarly if the condition (\ref{A.9}) had been chosen to define the Principal transform we would have:

\begin{equation}
\label{A.16}
\partial_i\hat h^i_j-\frac{1}{2}\partial_j\hat h
=-\frac{1}{2}\partial_jh_{00}  
\end{equation}
and this requires $\lambda$ to be:

\begin{equation}
\label{A.18}
\lambda=1
\end{equation}
in (\ref{1.13.9}).

Notice however that while either condition (\ref{A.15}) or (\ref{A.16}) is compatible with (\ref{1.13.9}) if $\hat R=0$ and therefore can be required for $\hat h^{\leftmoon}$ and $\hat h^{\sun}$, this is not true for $\hat h^{\earth}$ and in this case particular values of $C$ and $K$ in (\ref{2.5}) have to be chosen. Considering that the center of the Earth could be modeled by a perfect fluid the quo-harmonic condition requires $K=(1/3)\rho_0=0$, where $\rho_0$ is the central density and therefore this condition is here unacceptable. On the contrary the harmonic condition requires $K+2C=p_0=0$ where $p_0$ is the central pressure and the condition is acceptable if it is interpreted  as saying that $p_0<<\rho_0$.  


\section{Acknowledgements} 

I gratefully acknowledge the hospitality of the Theoretical Physics Department of the UPV/EHU as well as comments and TeX help from A. Molina.

\end{document}